\documentclass[12pt]{article}

\catcode`@=12
\begin{document}
\thispagestyle{empty}
\begin{flushright} 
UCRHEP-T328\\ 
January 2002\
\end{flushright}
\vspace{0.5in}
\begin{center}
{\LARGE	\bf New U(1) Gauge Extension of the\\
Supersymmetric Standard Model\\}
\vspace{1.5in}
{\bf Ernest Ma\\}
\vspace{0.2in}
{Physics Department, University of California, Riverside, 
California 92521, USA\\}
\vspace{1.5in}
\end{center}
\begin{abstract}\
In extending the minimal standard model of quarks and leptons to include 
supersymmetry, the conservation of baryon and lepton numbers is no longer 
automatic.  I show how the latter may be achieved with a new U(1) gauge 
symmetry and new supermultiplets at the TeV scale.  Neutrino masses and 
a solution of the $\mu$ problem are essential features of this proposed 
extension.
\end{abstract}

\newpage
\baselineskip 24pt

It is well-known that the minimal standard model of quarks and leptons 
conserves both baryon number $B$ and lepton number $L$ automatically 
(as the consequence of the assumed $SU(3)_C \times SU(2)_L \times U(1)_Y$ 
gauge symmetry and its representation content).  It is also well-known 
that this is not true any more once it is extended to include 
supersymmetry.  Thus any such extension must be supplemented by a new 
symmetry which forbids the violation of $B$ or $L$ or both.  There are 
many ways to do this; the most direct is to impose the conservation of 
an odd-even discrete symmetry, i.e. $R \equiv (-1)^{2j+3B+L}$, which is 
of course the defining hypothesis of the Minimal Supersymmetric 
Standard Model (MSSM).

There are two additional features of the MSSM which are often called into 
question.  One is the absence of neutrino masses.  This is, however, 
easily remedied by the addition of three neutral singlet lepton superfields 
(analogs of the three right-handed singlet neutrinos of the nonsupersymmetric 
standard model).  The other is the presence of the so-called $\mu$ term in the 
MSSM superpotential, i.e. $\mu \hat \phi_1 \hat \phi_2$, where $\hat 
\phi_{1,2}$ are the two Higgs superfields which spontaneously break the 
electroweak gauge symmetry.  Since this term is allowed by the gauge symmetry 
and the supersymmetry, there is no understanding of why $\mu$ should be the 
order of the electroweak breaking scale, rather than some very large 
unification scale.

Whereas there are piecemeal solutions of all the above three problems of the 
MSSM, it is clearly desirable to have a single principle which works for all 
three at the same time.  In this paper I show how a new simple U(1) gauge 
extension of the MSSM may be used exactly for this purpose \cite{prev}.

Consider the gauge group $SU(3)_C \times SU(2)_L \times U(1)_Y \times U(1)_X$. 
The usual quark and lepton (left-handed) chiral superfields transform as 
follows:
\begin{eqnarray}
&& (\hat u, \hat d) \sim (3,2,1/6;n_1), ~~ \hat u^c \sim (3^*,1,-2/3;n_2), 
~~ \hat d^c \sim (3^*,1,1/3;n_3), \\ 
&& (\hat \nu, \hat e) \sim (1,2,-1/2;n_4), ~~ \hat e^c \sim (1,1,1;n_5), 
~~ \hat N^c \sim (1,1,0;n_6).
\end{eqnarray}
They are supplemented by the two Higgs doublet superfields
\begin{equation}
\hat \phi_1 \sim (1,2,-1/2;-n_1-n_3), ~~~ \hat \phi_2 \sim (1,2,1/2;-n_1-n_2),
\end{equation}
with
\begin{equation}
n_1 + n_3 = n_4 + n_5, ~~~ n_1 + n_2 = n_4 + n_6,
\end{equation}
as in the MSSM.  However, the $\mu$ term is replaced by the trilinear 
interaction $\hat \chi \hat \phi_1 \hat \phi_2$, where $\hat \chi$ is a 
Higgs singlet superfield transforming as
\begin{equation}
\hat \chi \sim (1,1,0;2n_1+n_2+n_3).
\end{equation}
Thus
\begin{equation}
2n_1 + n_2 + n_3 \neq 0
\end{equation}
is required so that the effective $\mu$ parameter of this model is determined 
by the $U(1)_X$ breaking scale, i.e. $\langle \hat \chi \rangle$.

To complete this model, I add two copies of the singlet quark superfields
\begin{equation}
\hat U \sim (3,1,2/3;n_7), ~~~ \hat U^c \sim (3^*,1,-2/3;n_8),
\end{equation}
and one copy of
\begin{equation}
\hat D \sim (3,1,-1/3;n_7), ~~~ \hat D^c \sim (3^*,1,1/3;n_8),
\end{equation}
with
\begin{equation}
n_7 + n_8 = -2n_1 - n_2 - n_3,
\end{equation}
so that their masses are also determined by the $U(1)_X$ breaking scale.

To ensure the absence of the axial-vector anomaly \cite{ava}, the following 
conditions are considered \cite{gema}. 
\begin{equation}
[SU(3)]^2 U(1)_X ~:~ 2n_1 + n_2 + n_3 + n_7 + n_8 = 0,
\end{equation}
\begin{eqnarray} 
[SU(2)]^2 U(1)_X ~:~ 3 (3n_1 + n_4) + (-n_1-n_3) + (-n_1-n_2) = 0,
\end{eqnarray}
\begin{eqnarray}
[U(1)_Y]^2 U(1)_X &:& 3 \left[ 6 \left( {1 \over 6} \right)^2 n_1 + 3 \left( 
-{2 \over 3} \right)^2 n_2 + 3 \left( {1 \over 3} \right)^2 n_3 
+ 2 \left( -{1 \over 2} \right)^2 n_4 + n_5 \right] \nonumber \\ 
&& + 2 \left[ 3 \left( {2 \over 3} \right)^2 n_7 + 3 \left( -{2 \over 3} 
\right)^2 n_8 \right] + 3 \left( -{1 \over 3} \right)^2 n_7 + 3 \left( 
{1 \over 3} \right)^2 n_8 \nonumber \\ && + 2 \left( -{1 \over 2} \right)^2 
(-n_1-n_3) + 2 \left( {1 \over 2} \right)^2 (-n_1-n_2) = 0,
\end{eqnarray}
\begin{eqnarray} 
U(1)_Y [U(1)_X]^2 &:& 3 \left[ 6 \left( {1 \over 6} \right) n_1^2 + 3 \left( 
-{2 \over 3} \right) n_2^2 + 3 \left( {1 \over 3} \right) n_3^2 
+ 2 \left( -{1 \over 2} \right) n_4^2 + n_5^2 \right] \nonumber \\ 
&& + 2 \left[ 3 \left( {2 \over 3} \right) n_7^2 + 3 \left( -{2 \over 3} 
\right) n_8^2 \right] + 3 \left( -{1 \over 3} \right) n_7^2 + 3 \left( 
{1 \over 3} \right) n_8^2 \nonumber \\ 
&& + 2 \left( -{1 \over 2} \right) (-n_1-n_3)^2 + 2 \left( {1 \over 2} 
\right) (-n_1-n_2)^2 = 0,
\end{eqnarray}
\begin{eqnarray}
[U(1)_X]^3 &:& 3 \left[ 6 n_1^3 + 3 n_2^3 + 3 n_3^3 + 2 n_4^3 + n_5^3 + n_6^3 
\right] + 3 (3n_7^3 + 3 n_8^3) \nonumber \\ && + 2 (-n_1-n_3)^3 + 
2 (-n_1-n_2)^3 + (2n_1+n_2+n_3)^3 = 0.
\end{eqnarray}

Using Eq.~(9), it is clear that Eq.~(10) is automatically satisfied.  Using 
Eqs.~(4) and (9), it is easily shown that both Eqs.~(11) and (12) are 
satisfied by the single condition
\begin{equation}
n_2 + n_3 = 7 n_1 + 3 n_4.
\end{equation}
Using Eqs.~(4), (9), and (15), it is then simple to show that Eq.~(13) becomes
\begin{equation}
6(3n_1+n_4)(2n_1-4n_2-3n_7) = 0.
\end{equation}
Using Eq.~(15), it is clear that $3n_1+n_4 = 0$ contradicts Eq.~(6).  Hence 
only the condition
\begin{equation}
2n_1 - 4n_2 - 3n_7 = 0
\end{equation}
will be considered from here on.

At this point, the eight parameters ($n_1$ to $n_8$) are constrained by the 
five conditions given by Eqs.~(4), (9), (15), and (17).  Consider $n_1$, 
$n_4$, and $n_6$ as the independent parameters.  The others are then given by
\begin{eqnarray}
n_2 &=& -n_1 + n_4 + n_6, \\ 
n_3 &=& 8n_1 + 2n_4 - n_6, \\ 
n_5 &=& 9n_1 + n_4 - n_6, \\ 
n_7 &=& 2n_1 - {4 \over 3} n_4 - {4 \over 3} n_6, \\ 
n_8 &=& -11 n_1 - {5 \over 3} n_4 + {4 \over 3} n_6.
\end{eqnarray}
It is now straightforward to simplify Eq.~(14) to read
\begin{equation}
-36 (3n_1+n_4)(9n_1+n_4-2n_6)(6n_1-n_4-n_6) = 0.
\end{equation}
Whereas one factor, i.e. $3n_1+n_4$, must be nonzero, there remain two 
possible solutions, i.e.
\begin{eqnarray}
({\rm A}) && n_6 = {1 \over 2} (9n_1+n_4), \\ 
({\rm B}) && n_6 = 6n_1 - n_4,
\end{eqnarray}
which render $U(1)_X$ free of the axial-vector anomaly.  This exact factoring 
of the sum of eleven cubic terms is certainly not a trivial result \cite{eg}.

Solution (A) is thus given by
\begin{eqnarray}
&& n_2 = n_3 = {1 \over 2} (7n_1+3n_4), ~~~ n_5 = n_6 = {1 \over 2} 
(9n_1 + n_4), \\ && n_7 = -4n_1 - 2n_4, ~~~ n_8 = -5n_1 - n_4.
\end{eqnarray}
In the MSSM, $\hat L$ and $\hat \phi_1$ transform identically under 
$SU(3)_C \times SU(2)_L \times U(1)_Y$.  Here $\hat L$ and $\hat \phi_1$ are 
distinguished by $U(1)_X$ if
\begin{equation}
9n_1 + 5n_4 \neq 0.
\end{equation}
Hence the lepton number $L$ may be automatically conserved as in the 
nonsupersymmetric standard model.

In the MSSM, the term $\hat u^c \hat d^c \hat d^c$ is allowed in the 
superpotential.  Here it is forbidden if
\begin{equation}
7n_1 + 3n_4 \neq 0.
\end{equation}
Hence the baryon number $B$ may be automatically conserved as well.

Solution (B) has
\begin{eqnarray}
&& n_2 = 5n_1, ~~ n_3 = 2n_1 + 3n_4, ~~ n_5 = 3n_1 + 2n_4, \\ 
&& n_6 = 6n_1 - n_4, ~~ n_7 = -6n_1, ~~ n_8 = -3n_1 - 3n_4.
\end{eqnarray}
Hence $L$ is automatically conserved if
\begin{equation}
3n_1 + 4n_4 \neq 0,
\end{equation}
and $B$ is automatically conserved if 
\begin{equation}
3n_1 + 2n_4 \neq 0.
\end{equation}
Note that solutions (A) and (B) are identical if $n_4 = n_1$.  This turns 
out to be also the condition \cite{coc} for $U(1)_X$ to be orthogonal to 
$U(1)_Y$, i.e.
\begin{eqnarray} 
&& 3 \left[ 6 \left( {1 \over 6} \right) n_1 + 3 \left( 
-{2 \over 3} \right) n_2 + 3 \left( {1 \over 3} \right) n_3 
+ 2 \left( -{1 \over 2} \right) n_4 + n_5 \right] \nonumber \\ 
&& + 2 \left[ 3 \left( {2 \over 3} \right) n_7 + 3 \left( -{2 \over 3} 
\right) n_8 \right] + 3 \left( -{1 \over 3} \right) n_7 + 3 \left( 
{1 \over 3} \right) n_8 \nonumber \\ 
&& + 2 \left( -{1 \over 2} \right) (-n_1-n_3) + 2 \left( {1 \over 2} 
\right) (-n_1-n_2) = 0.
\end{eqnarray}

There are two more anomalies to consider.  The global SU(2) chiral gauge 
anomaly \cite{witten} is absent because the number of $SU(2)_L$ doublets 
is even.  The mixed gravitational-gauge anomaly \cite{mixed} is proportional 
to the sum of $U(1)_X$ charges, i.e.
\begin{eqnarray}
&& 3(6n_1 + 3n_2 + 3n_3 + 2n_4 + n_5 + n_6) + 3(3n_7 + 3n_8) \nonumber \\ 
&& + 2(-n_1-n_3) + 2(-n_1-n_2) + (2n_1 + n_2 + n_3) = 6(3n_1 + n_4),
\end{eqnarray}
which is not zero.  This anomaly may be tolerated if gravity is neglected. 
On the other hand, it may be rendered zero by adding $U(1)_X$ 
supermultiplets as follows: one with charge $3(3n_1+n_4)$, three with charge 
$-2(3n_1+n_4)$, and three with charge $-(3n_1+n_4)$.  Hence they contribute 
$3 + 3(-2-1) = -6$ (in units of $3n_1+n_4$) to Eq.~(35), but $27 + 3(-8-1) 
= 0$ to Eq.~(14).

The allowed terms in the superpotential of either solution (A) or (B) 
consist of the usual allowed terms of the MSSM with $\mu \hat \phi_1 
\hat \phi_2$ replaced by $\hat \chi \hat \phi_1 \hat \phi_2$.  In (A), 
the usual $R$-parity violating terms are forbidden by Eqs.~(28) and (29). 
As for the interactions of the exotic quark singlets of Eqs.~(7) and (8), 
$n_1 + n_4 \neq 0$ forbids $\hat U^c \hat d^c \hat d^c$, $\hat u^c \hat d^c 
\hat D^c$, and $(\hat u_i \hat d_j - \hat d_i \hat u_j) \hat D$; 
$13n_1 + n_4 \neq 0$ forbids $\hat U^c \hat d^c \hat D^c$; and $n_1 \neq 0$ 
forbids $\hat e^c \hat u^c \hat D$, $(\hat \nu \hat d - \hat e \hat u) 
\hat D^c$, $\hat N^c \hat u^c \hat U$, and $\hat N^c \hat d^c \hat D$. 
This means that if $n_4 = -n_1$, then $\hat U^c$ and $\hat D^c$ are diquark 
superfields, and if $n_1 = 0$, then $\hat U$ and $\hat D$ are leptoquark 
superfields.

In solution (B), the usual $R$-parity violating terms are forbidden by 
Eqs.~(32) and (33).  Furthermore, $n_1+3n_4 \neq 0$ forbids $\hat U^c 
\hat d^c \hat d^c$; $n_1 \neq 0$ forbids $\hat u^c \hat d^c \hat D^c$ and 
$(\hat u_i \hat d_j - \hat d_i \hat u_j) \hat D$; $4n_1 + 3n_4 \neq 0$ 
forbids $\hat U^c \hat d^c \hat D^c$; $n_1+n_4 \neq 0$ forbids $\hat e^c 
\hat u^c \hat D$, $(\hat \nu \hat d - \hat e \hat u) \hat D^c$, and 
$\hat N^c \hat d^c \hat D$; and $5n_1-n_4 \neq 0$ forbids $\hat N^c 
\hat u^c \hat U$.  This means that $\hat U^c$ is a diquark if $n_4 = -n_1/3$, 
and $\hat U$ is a leptoquark if $n_4 = 5n_1$; whereas $\hat D^c$ is a 
diquark if $n_1 = 0$, and $\hat D$ is a leptoquark if $n_4 = -n_1$.

Even with the imposition of $R$-parity, there are higher-dimensional operators 
in the MSSM which may induce proton decay, i.e. $\hat q \hat q \hat q 
\hat l$ and $\hat u^c \hat u^c \hat d^c \hat e^c$.  In the nonsupersymmetric 
standard model, since quarks and leptons are fermions, these operators have 
dimension six, but in the MSSM they have dimension-five pieces.  Hence 
proton decay may not be sufficiently suppressed, which is a well-known 
problem of the MSSM.  In this model, these terms are forbidden [in both 
solutions (A) and (B)] by $3n_1+n_4 \neq 0$, i.e. the same condition that 
forbids the $\mu$ term.

As it stands, this model pairs $\nu$ with $N^c$ to form a Dirac neutrino 
with mass proportional to $\langle \phi_2 \rangle$.  This would require 
extremely small Yukawa couplings and is generally considered to be very 
unnatural.  On the other hand, if $n_6 = 3n_1 + n_4$ is assumed [i.e. 
$n_4 = 3n_1$ in solution (A) or $n_4 = 3n_1/2$ in solution (B)], then the 
extra singlets used to cancel the mixed gravitational-gauge anomaly of 
Eq.~(35) are exactly of the right number and structure to allow neutrinos 
to acquire naturally small seesaw Dirac masses, as shown below.

In addition to the 3 singlets $N^c$ of $U(1)_X$ charge $n_6$, there are now 
also 3 singlets $N$ of charge $-n_6$ and 3 singlets $S^c$ of charge $-2n_6$. 
The $12 \times 12$ mass matrix spanning $(\nu, S^c, N, N^c)$ is then of the 
form
\begin{equation}
{\cal M} = \left[ \begin{array} {c@{\quad}c@{\quad}c@{\quad}c} 0 & 0 & 0 & m_1 
\\ 0 & 0 & m_2 & 0 \\ 0 & m_2 & 0 & M \\ m_1 & 0 & M & 0 \end{array} \right],
\end{equation}
where $m_1$ comes from $\nu N^c \phi_2^0$ with $\langle \phi_2^0 \rangle 
\neq 0$, $m_2$ comes from $N S^c \chi$ with $\langle \chi \rangle \neq 0$, 
and $M$ is an allowed invaraint mass.  Thus $m_1 \sim 10^2$ GeV, $m_2 \sim 
10^3$ GeV, and $M \sim 10^{16}$ GeV are expected.  In the reduced $(\nu, S^c)$ 
sector, the effective $6 \times 6$ mass matrix is still exactly of the Dirac 
form, i.e.
\begin{equation}
{\cal M}_\nu = \left[ \begin{array} {c@{\quad}c} 0 & -m_1 m_2/M \\ -m_1 m_2/M 
& 0 \end{array} \right],
\end{equation}
and $m_1 m_2/M \sim 10^{-2}$ eV is the right order of magnitude for 
realistic neutrino masses.

In conclusion, a remarkable new U(1) gauge symmetry has been identified in a 
simple extension of the supersymmetric standard model which is capable of 
enforcing $B$ or $L$ conservation or both, as well as the absence of the $\mu$ 
term and the presence of neutrino masses.  Two solutions have been obtained 
[from the exact factoring of Eq.~(14) to become Eq.~(23)] with many possible 
variations regarding new interactions beyond the MSSM, as summarized in 
Tables 1 and 2.  The origin of this new U(1) gauge symmetry is unknown at 
present; it has no obvious fit into any simple model of grand unification 
or string theory.

This work was supported in part by the U.~S.~Department of Energy under 
Grant No.~DE-FG03-94ER40837.

\begin{table}[htb]
\caption{Solutions (A) and (B) where $n_i = a n_1 + b n_4$.}
\begin{center}
\begin{tabular}{||c|c|c||c|c||}
\hline \hline
& \multicolumn{2}{c||}{(A)} & \multicolumn{2}{c||}{(B)} \\ 
\cline{2-5}
 & $a$ & $b$ & $a$ & $b$ \\ 
\hline
$n_2$ & 7/2 & 3/2 & 5 & 0 \\ 
$n_3$ & 7/2 & 3/2 & 2 & 3 \\ 
$n_5$ & 9/2 & 1/2 & 3 & 2 \\ 
$n_6$ & 9/2 & 1/2 & 6 & --1 \\ 
$n_7$ & --4 & --2 & --6 & 0 \\ 
$n_8$ & --5 & --1 & --3 & --3 \\ 
\hline
$-n_1-n_3$ & --9/2 & --3/2 & --3 & --3 \\ 
$-n_1-n_2$ & --9/2 & --3/2 & --6 & 0 \\ 
$2n_1+n_2+n_3$ & 9 & 3 & 9 & 3 \\ 
\hline \hline
\end{tabular}
\end{center}
\end{table}

\begin{table}[htb]
\caption{Conditions on $n_1$ and $n_4$ in (A) and (B).}
\begin{center}
\begin{tabular}{||c|c||c|c|l||}
\hline \hline
\multicolumn{2}{||c||}{(A)} & \multicolumn{2} {c|}{(B)} & \\ 
\hline
$c$ & $d$ & $c$ & $d$ & $cn_1+dn_4 \neq 0$ forbids \\ 
\hline
3 & 1 & 3 & 1 & $\mu$ term \\ 
9 & 5 & 3 & 4 & $L$ violation \\ 
7 & 3 & 3 & 2 & $B$ violation \\ 
1 & 1 & 1 & 3 & $U^c$ as diquark \\ 
1 & 1 & 1 & 0 & $D^c$ as diquark \\ 
1 & 0 & 5 & --1 & $U$ as leptoquark \\ 
1 & 0 & 1 & 1 & $D$ as leptoquark \\ 
13 & 1 & 4 & 3 & $U^c$, $D^c$ as semiquarks \\ 
\hline \hline
\end{tabular}
\end{center}
\end{table}

\bibliographystyle{unsrt}

\begin{thebibliography}{99}
\bibitem{eg} Consider for example $n_1=2/3$, $n_4=-1$ in solution (A), then 
Eq.~(14) is realized as $18(2/3)^3 + 18(5/6)^3 + 6(-1)^3 + 6(5/2)^3 + 
9(-2/3)^3 + 9(-7/3)^3 + 4(-3/2)^3 + (3)^3 = 0$.
\end{thebibliography}

\end{document}